\documentclass[%
 reprint,
 superscriptaddress,
 amsmath,amssymb,
 aps,
floatfix,
]{revtex4-2}

\usepackage{graphicx}% Include figure files
\usepackage{dcolumn}% Align table columns on decimal point
\usepackage{bm}% bold math
\usepackage{array}% table things
\usepackage[hidelinks]{hyperref}% add hypertext capabilities
\usepackage{graphicx, nicefrac} 
\usepackage{comment} % to add comments
\usepackage{threeparttable} % for more complex tables
\usepackage{tabularx} % in the preamble
\usepackage{wasysym} % for the diameter symbol
\usepackage{float}
\usepackage[
separate-uncertainty = true,
multi-part-units=single
]{siunitx}
\usepackage{xcolor}
\usepackage[version=4]{mhchem}
\usepackage[normalem]{ulem}
\usepackage{tablefootnote}
\usepackage{upgreek} % for greek letters that aren't variables, e.g. $\upmu\mathrm{s}$

\begin{document}

%\preprint{APS/123-QED}

\title{Reflection of a Diffuser in a Liquid Interface}
% 289 
\author{C.~Silva}
\email{Corresponding author: claudio.silva@coimbra.lip.pt}
\affiliation{{Laborat\'orio de Instrumenta\c c\~ao e F\'isica Experimental de Part\'iculas (LIP)}, University of Coimbra, P-3004 516 Coimbra, Portugal}

% 52 
\author{R.~Cabrita}
\affiliation{{Laborat\'orio de Instrumenta\c c\~ao e F\'isica Experimental de Part\'iculas (LIP)}, University of Coimbra, P-3004 516 Coimbra, Portugal}

% 295 
\author{V.N.~Solovov}
\affiliation{{Laborat\'orio de Instrumenta\c c\~ao e F\'isica Experimental de Part\'iculas (LIP)}, University of Coimbra, P-3004 516 Coimbra, Portugal}

% 46 
\author{P.~Br\'{a}s}
\affiliation{{Laborat\'orio de Instrumenta\c c\~ao e F\'isica Experimental de Part\'iculas (LIP)}, University of Coimbra, P-3004 516 Coimbra, Portugal}

% 179 
\author{A.~Lindote}
\affiliation{{Laborat\'orio de Instrumenta\c c\~ao e F\'isica Experimental de Part\'iculas (LIP)}, University of Coimbra, P-3004 516 Coimbra, Portugal}

% 252 
\author{G.~Pereira}
\affiliation{{Laborat\'orio de Instrumenta\c c\~ao e F\'isica Experimental de Part\'iculas (LIP)}, University of Coimbra, P-3004 516 Coimbra, Portugal}

% 187 
\author{M.I.~Lopes}
\affiliation{{Laborat\'orio de Instrumenta\c c\~ao e F\'isica Experimental de Part\'iculas (LIP)}, University of Coimbra, P-3004 516 Coimbra, Portugal}

\date{\today}% It is always \today, today,

%Use showkeys class option if keyword

\begin{abstract}

We present a novel method, based on the Saunderson corrections, to predict the reflectance between a liquid interface and a dielectric diffuser. In this method, the diffuse properties of the dielectric are characterized using a single parameter, the multiple-scattering albedo, which is the same irrespective of being in contact with air or liquid. We tested this method using an apparatus based on a total integrating sphere capable of measuring reflectance in both liquid and gas interfaces across various wavelengths of light. We observed that the difference in the value of the multiple-scattering albedo between the sphere full of liquid and empty was less than 0.9$\times 10^{-3}$, with the average difference normalized to the respective uncertainty of only 0.7. These results confirm the reliability of our method and its potential for use in a wide range of practical applications.
\end{abstract}

\newcommand{\CICLO}{C$_{\mathrm{6}}$H$_{\mathrm{12}}$}

\maketitle

\section{Introduction}{The description of the reflectance within a liquid medium is needed in many physics applications such as liquid scintillators or Cherenkov detectors and computer vision, the computer rendering of realistic images. However, obtaining accurate measurements of the optical properties of surfaces submerged in a liquid presents significant challenges compared to those in air, and as such, measurements in a liquid interface are not typically available, especially for diffuse dielectric reflectors. Diffuse reflection occurs when the light is refracted to the bulk of an inhomogeneous dielectric material. The inhomogeneities act as scatter centers in an otherwise uniform dielectric medium, causing the light to scatter multiple times before returning to the first medium. 

Most of the diffuse materials look darker when wet. Two main explanations have been proposed to explain this phenomenon: i) the penetration of the liquid into porous materials reduces the contrast between the refractive index of the pore and the material, increasing the forward scattering and thus increasing the probability of absorption \cite{Twomey:86}; ii) internal reflection in the liquid layer covering the surface increases the likelihood of absorption by the surface \cite{Lekner:88}. Nonetheless,  observations made directly in the liquid medium show an increase in reflectance. For example, Voss and Zhang observed that the reflectance of a plate made of Spectralon\textregistered~increases by 2\% when that plate is submerged in water \cite{Voss:06}. Also, in particle physics detectors, which is of particular interest to us, the reflectance of polytetrafluoroethylene (PTFE) to the 178~nm xenon excimer emission light \cite{Jortner1965LocalizedEI} increases from $\sim$75\% to 95\% when immersed in liquid xenon \cite{Neves:2016tcw, 10.1117/12.344390}. It should be noted that the temperature of the liquid might also affect the reflectance in the latter case, which is of particular interest in particle physics detectors.

The reflection at a liquid interface is critical in designing particle detectors since many applications in this field use liquids as detection media. Examples include water Cherenkov and scintillator detectors \cite{FUKUDA2003418, BOGER2000172}, organic scintillators \cite{PhysRevLett.100.221803}, or more recently, liquefied noble gases such as xenon, argon, and helium \cite{AKERIB2020163047, PhysRevD.98.102006}. In the case of scintillation detectors, the observed optical signal is proportional to the deposited energy, making the internal reflectance an essential parameter for detector performance.  For applications like dark matter or coherent elastic neutrino-nucleus scattering (CE$\nu$NS) \cite{RevModPhys.82.2053}, maximizing light collection is crucial to decrease the energy threshold and enable detection. Typically, these detectors use PTFE as an efficient reflector material, and simulating the optical properties of the light collection model requires reflectance properties of surfaces as input. However, standard reflectometers measure reflectance only in air, and measuring reflectance in liquid is complex due to uncertainties arising from light absorption, scattering, and bubble formation. Therefore, predicting reflectance in liquid based on values observed in gas can help overcome the challenges associated with liquid reflectance measurements.

This article is structured as follows: first, we present the method to describe the reflectance of a diffuser, irrespective of the interface. This method is based on the previous work of Lawrence Wolff \cite{Wolff:94} and uses the Saunderson model \cite{Saunderson:42} in which the internal reflections between the diffuser and the original medium, which might be air or the liquid, are considered directly (sec.\,\ref{SecModel}). In this method, the optical properties of the diffuser are described using a single parameter, the multiple-scattering albedo, $\rho$, that does not depend directly on the first medium. To test these assumptions, we built a setup based on a total integrating sphere that can be filled with different liquids (sec.\,\ref{sec:Setup}). Then, we implemented this model and the geometry of the setup in a Monte Carlo simulation based on the ANTS2 software package \cite{ANTS} (sec.\,\ref{sec:Simulations}) and compared these results with the equation of the sphere derived for our specific geometry. Using these simulations, we obtained the value of the throughput of the sphere for different values of the single scatter albedo. The results are presented in the sec.\, \ref{sec:Results}. Finally, in sec.\, \ref{sec:Discussion}, we discuss these results and present an analytical method to obtain the multiple-scattering albedo when the hemispherical reflectance of a surface is known and discuss possible expansions of the current model.

\section{Modeling the Diffuse Reflection}\label{SecModel}

Consider two dielectric media in optical contact. One of these media, further designated as \emph{Medium 1}, is optically transparent with refractive index $n_1$,  while another, which we call \emph{diffuser} is optically inhomogeneous, meaning that the refractive index varies from place to place in the dielectric volume. We also assume that the diffuser is semi-infinite, meaning that it is thick enough that no light is transmitted through the material. The refracted light scatters multiple times in these inhomogeneities before being absorbed or returning to the boundary between the diffuser and the medium $1$ (internal boundary). If the light returns to the internal boundary, it can be reflected back to the diffuser (internal reflection) or be refracted to medium $1$. If it is refracted, it is part of the diffuse lobe. In the case of internal reflection, the light undergoes a multiple-scatter process in the diffuser again. This process continues until all the light is either absorbed or returns to the medium $1$. The internal scattering process depends only on the material's optical characteristics, but the refractions into and from the diffuser and the internal reflections must obey the Fresnel equations \cite{Born:382152}, adding a dependence on the optical properties of the first medium as well.

In the description of the diffuse reflectance, the most common approach is to model the reflectance of these surfaces using the Lambert law, which states that diffuse materials appear equally bright independently of the viewing angle; therefore, the bidirectional reflectance intensity distribution function, $f_r$ \cite{nicodemus}, of the surface is given by a constant, ($f_r = \rho_l/\pi$), usually called albedo of the surface. However, as discussed, both refractions add deviations to this law, especially for surfaces illuminated or observed at a large angle. Therefore, L.~Wolff modified the Lambert law by introducing two factors accounting for the light that is reflected at both the entrance and exit of the diffuser \cite{Wolff:94}. In his model, $f_r$ is given by the following sum of two components, specular and diffuse:
\begin{align}
\begin{split}
f_r\left(\theta_i, \theta_r, \phi_i, \phi_r \right) & = F\left(\theta_i; \frac{n_2}{n_1},\right)\delta\left(\theta_i - \theta_r\right)\delta\left(\phi_i + \phi_r\right) + \\
 & + \frac{1}{\pi}\varrho_d \left[1-F\left(\theta_i; \frac{n_2}{n_1}\right)\right] \times \\ & \times \left[1-F\left(\arcsin\left(\frac{n_1}{n_2}\sin\theta_r\right); \frac{n_1}{n_2}\right)\right],
\label{BRIDF_mymodel}
\end{split}
  \end{align}
 where the angles $\theta_i$ and $\theta_r$ are the polar angles of the incident direction of the light (subscript $i$) and the viewing direction (subscript $r$), $\phi_i$ and $\phi_r$ the corresponding azimuthal angles, $n_2$ the average refractive index of the diffuser, and $\varrho_d$ the total diffuse albedo. $F\left(\theta; n, \alpha\right)$ corresponds to the reflectivity calculated by the Fresnel Equations (see the appx.\,\ref{AppendixA}, eq.\,\ref{Eq:Fresnel}). Since the light is partially polarized when refracted to the diffuser but effectively depolarized after the multiple-scattering process, the polarization of the light is not considered in the Fresnel equations.

In our earlier work \cite{ReflectanceJAPArticle}, we demonstrated that our model accurately reproduces the distribution of reflected light from a diffuser in air, such as PTFE. However, this model has two main limitations. First, the second Fresnel factor (eq.~\ref{BRIDF_mymodel}) decreases the reflectance along a particular direction, but this light is not absorbed, instead being reflected in another direction. This increase in the reflectance is accounted for in the total diffuse albedo $\varrho_d$, but in highly reflective surfaces, $\varrho_d$ is often larger than 1, which may seem counterintuitive. Second, $\varrho_d$ depends not only on the optical properties of the diffuse medium but also on its refractive index of the medium $1$ since it includes all the additional reflections at the internal boundary (as shown in eq.\,4 in ref. \cite{Wolff:94}). Therefore, if the medium $1$ changes $\varrho_d$ needs to be estimated again. 

To address these two issues, we have introduced a new albedo, the multiple-scattering albedo $\rho$, which replaces the total diffuse albedo in the eq.~\ref{BRIDF_mymodel}. $\rho$ represents the probability that the light refracted to the diffuser or reflected in the internal boundary is not absorbed in the multiple scattering and returns to the boundary between the diffuser and medium $1$. This approach is similar to Saunderson's method for describing the reflectance of pigment plastics at an air interface in 1942 \cite{Saunderson:42}, which utilized the Kubelka and Munk theory \cite{Kubelka}. Our model assumes two things: first, that $\rho$ is independent of the angle of incidence, and second, that the direction of light is random after multiple scattering, which means that light should follow Lambert's law before being refracted or reflected.

Since the light can be reflected back to the diffuser multiple times, the multiple-scattering albedo relates with the total diffuse albedo through the following summation:
\begin{equation}
\begin{split}
\varrho_d = & \rho + \overline{\mathcal{F}}_{\nicefrac{n_1}{n_2}} \rho + \overline{\mathcal{F}}_{\nicefrac{n_1}{n_2}}^2 \rho^2 + .... \\
 = & \rho\left(1-\rho\overline{\mathcal{\mathcal{F}}}_{\nicefrac{n_1}{n_2}}\right)^{-1},
 \label{eq:multiplescattering}
\end{split}
\end{equation}
where $\overline{\mathcal{F}}_{\nicefrac{n_1}{n_2}}$ corresponds to the probability of reflection between an interface of refractive index $n_2$ and an interface of refractive index $n_1$. Assuming that the photons arriving at the surface follow Lambertian law, it is given by the integral:
\begin{equation}
\overline{\mathcal{F}}_{\nicefrac{n_1}{n_2}} = \frac{1}{\pi}\int_{2\pi} F \left(\theta; \frac{n_1}{n_2}\right) \mathrm{d}\Omega ,
\label{IntegralFo}
\end{equation}
where $\theta$ is the angle of reflection and  $\mathrm{d}\Omega$  the element projected solid angle  defined as
\begin{equation}
\mathrm{d}\Omega = \cos\theta \sin \phi  \mathrm{d}\theta \mathrm{d}\phi.
\label{eq:dOmega}
\end{equation}
This definition solves the two issues mentioned before: $\rho\leq$1 since it is directly linked to a probability, and it is independent of the optical properties of medium $1$ since that information is included in the factor $\overline{\mathcal{F}}_{\nicefrac{n_1}{n_2}}$. The equation \ref{IntegralFo} is integrable for an interface between two dielectrics. The result of the integral is presented in appx.\,\ref{AppendixA} eq.\,\ref{integralresultado}.}\label{sec:Introduction}

\section{Experimental Method}{To study the effect of the medium interface on surface reflectance, we build a setup aiming to measure the change in the throughput of a total integrating sphere when a liquid replaces the air volume. This setup, composed of four different experimental configurations, is represented in fig.\,\ref{Fig1_Setup}. In each configuration, its main elements are: a) the matrix of LEDs with wavelengths ranging from 255~nm until 490~nm, b) a system of collimation and beam sampling, c) the total integrating sphere (TIS), and d) the acquisition system composed by two photomultipliers (PMT) operating in photon counting mode. Using this setup, we can measure the observed flux, $\Phi_R$, after the light has been reflected in the sphere (fig.\,\ref{Fig1_Setup}A), and the incident flux, $\Phi_I$, entering the sphere (fig.\,\ref{Fig1_Setup}B). Furthermore, since the light reflected in the PMT can return back and be detected, the reflectance results depend on the reflectance of the PMT photocathode. As such, configurations C and D are used to measure the reflectance of the PMT photocathode.

The throughput of the sphere, $H$, is defined as the ratio between the observed reflected photon flux, $\Phi_R$, and the observed incident flux, $\Phi_I$:
\begin{equation}
H = \frac{\Phi_R}{\Phi_I},
\end{equation}
with both units measured in terms of the number of detected photons (phd) per unit of time.
To compare the effect of the medium interface on the sphere's throughput, we measure $H$ for both air and liquid interfaces and compare them with the results obtained from Monte Carlo simulations (sec.\,\ref{sec:Simulations}) for a given value of multiple-scattering albedo. The experimental method used to obtain $H_{\mathrm{air}}$ and $H_{\mathrm{liq}}$ is described below.

\subsection{The Experimental Set-Up}
\begin{figure}
\centering

\begin{comment}

\psfrag{LED}[c][c][1][1]{LED}
\psfrag{Matrix}[c][c][1][1]{Matrix}
\psfrag{Diffuser}[c][c][1][1]{Diffuser}
\psfrag{PMT}[c][c][1][1]{PMT}
\psfrag{ph}[c][c][1][1]{Pin-hole}
\psfrag{pinnhole}[c][c][1][1]{Pin-hole}
\psfrag{pinhhole}[c][c][1][1]{Pin-hole}

\psfrag{2 mm}[c][c][1][1]{2 mm}
\psfrag{1 mm}[c][c][1][1]{1 mm}
\psfrag{Beam}[c][c][1][1]{Beam}
\psfrag{Sampler}[c][c][1][1]{Sampler}
\psfrag{Quartz}[c][c][1][1]{Quartz}
\psfrag{Flat}[c][c][1][1]{Flat}
\psfrag{TIS}[c][c][1][1]{TIS}
\psfrag{Light}[c][c][1][1]{Light}
\psfrag{Trap}[c][c][1][1]{Trap}
\psfrag{Reference}[c][c][1][1]{Reference}
\psfrag{Main}[c][c][1][1]{Main}
\psfrag{Iris}[c][c][1][1]{Iris}
\psfrag{Incident Beam}[c][c][1][1]{Incident Beam}
\psfrag{Measurement}[c][c][1][1]{Measurement}
\psfrag{1.5 port}[c][c][1][1]{1.5 port}
\end{comment}

\includegraphics[width=8.4cm]{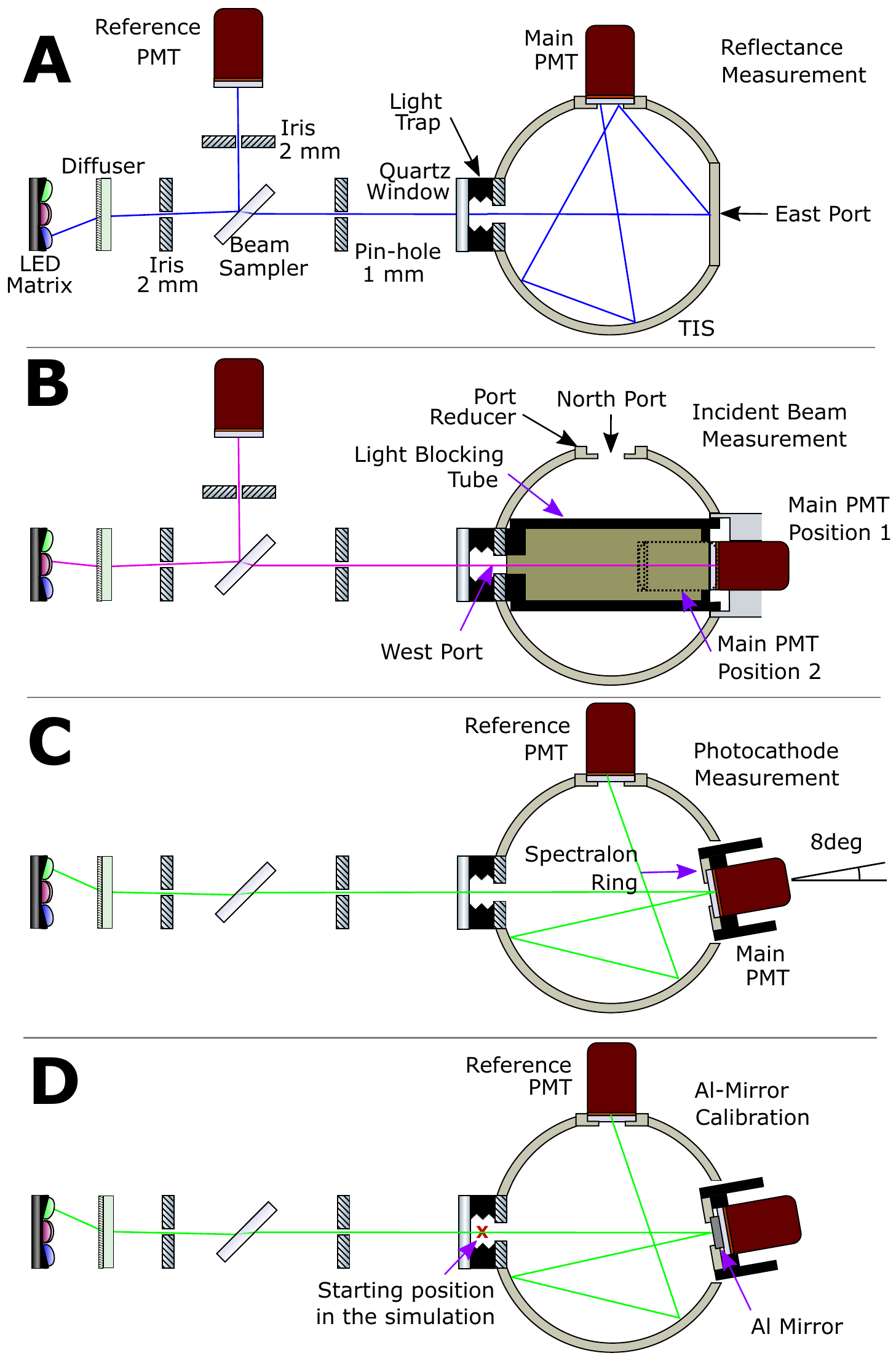}
\caption{Experimental optical set-up. Panel A: measurement of the reflectance with the 1.5\,in port closed and the main PMT mounted in the north port. Panel B: measurement of the incident beam with the main PMT mounted in the 1.5\,in port facing the incident beam at an angle of 0$^{\circ}$. Panel C: measurement of the PMT reflectance with the PMT mounted in the 1.5\,in port facing the incident beam at an angle of 8$^{\circ}$. Panel D: calibration of the PMT reflectance with the aluminium mirror placed in front of the PMT. The coloured lines represent a typical light ray path.}
\label{Fig1_Setup}
\end{figure}

\begin{table}[ht]
\caption{Characteristics of the LEDs used in the experimental set-up and the optical characteristics of fused silica and water for the specific LED wavelength}
\label{LEDTable}
%Here, $z$ is defined as vertical distance from the face of the bottom PMT array, accounting for thermal contraction as appropriate.
\begin{center}
%\begin{tabularx}{\linewidth}{@{}lYYYYYYY@{}}
\begin{threeparttable}
\begin{tabularx}{0.80\linewidth}{lScSSSS}
\hline
\hline
\multicolumn{1}{c}{$\lambda$} & FWHM & \multicolumn{1}{c}{n$_{\mathrm{SiO_{2}}}$\tnote{\dag}} & \multicolumn{1}{c}{n$_{\mathrm{H_{2}O}}$} &  
\multicolumn{1}{c}{$a_{\mathrm{min}}$\tnote{\ddag}} & \multicolumn{1}{c}{$a_{\mathrm{max}}$\tnote{\ddag}} 
\\
\multicolumn{1}{c}{[nm]} & \multicolumn{1}{c}{[nm]} & 
 & &
\multicolumn{1}{c}{[km$^{-1}$]}  &
\multicolumn{1}{c}{[km$^{-1}$]} \\
&&
\multicolumn{1}{c}{\cite{QuartzN}} & 
\multicolumn{1}{c}{\cite{WaterN}} & 
 \multicolumn{1}{c}{\cite{Absorption_Buiteveld, Absorption_Quickenden_and_Irvin}} &
\multicolumn{1}{c}{\cite{Absorption_Mason16}} \\

\hline
255 &11 &1.5048&1.3751  & 75.1& 51.50\\
275 &11  &1.4960&1.3668 & 49.7& 22.30 \\
285 &11  &1.4924&1.3634 & 42.6& 9.39 \\
310 &11  &1.4924&1.3568  & 25.8& 2.36  \\
356 &\multicolumn{1}{l}{10--20}  &1.4761&1.3488&  8.6& 0.98 \\
405 &19   &1.4696&1.3431 & 6.3& 2.48   \\
490 &30  &1.4629 &1.3373& 18.1& 14.60  \\
\hline
\hline
\end{tabularx}
\begin{tablenotes}
\item[\dag] Fused silica
\item[\ddag] Minimum and maximum value of the absorption coefficient of the water
\end{tablenotes}
\end{threeparttable}
\end{center}
\end{table}
% os valores de rayleigh scattering são do Irvin até 310 e do Buitevelt depois disso

\subsubsection{The Beam collimation}
The light is emitted from a set of 7 LEDs from Roithner Lasertechnik\textregistered, ranging from the UV ($\lambda$=255~nm) to the visible green ($\lambda$=490~nm). These LEDs exhibit a narrow spectral bandwidth of 10~nm FWHM (full width at half maximum) for the UV LEDs and between 20 and 30~nm FWHM for the visible LEDs. They are soldered in a 1-inch circular PCB and controlled via an electronic board plugged into an Arduino Uno\textregistered~microcontroller board. Further details on the characteristics and positioning of the LEDs can be found in tab.\,\ref{LEDTable}. 

The light emitted by the LEDs goes through a diffuser (DGUV10-220 from Thorlabs\textregistered) located 50~mm away from the matrix. The diffuser has a transmittance above 68\% for this selection of LEDs. Next, the light is collimated by a \diameter\,2~mm iris diaphragm before reaching the beam-sampler BSF10-UV from Thorlabs\textregistered, placed 209~mm from the matrix and at an angle of 45$^{\circ}$ relative to the direction of the incoming light. The beam sampler reflects the light with a probability between 0.5\% ($\lambda$=255~nm) and 2.8\% ($\lambda$=490~nm). The reflected light goes through a second \diameter\,2~mm iris diaphragm and is directed towards the reference photomultiplier (Hammamatsu\textregistered P762-Y001), located 86~mm away from the beam sampler. The transmitted light is further collimated using a \diameter\,1~mm pin-hole, placed 48~mm from the beam sampler. Then, it enters the total integrating sphere (TIS) through a 6~mm thick optical flat made of fused quartz from Crystran\textregistered. 

\subsubsection{The Total Integrating Sphere}

The total integrating sphere is the model 819C-SL-3.3 made of  Spectralon\textregistered\,(PTFE) from Newport\textregistered, with an internal diameter of 3.3~inches. The sphere has four ports, three of which have an aperture of 1~inch, and one with an aperture of 1.5~inches. Light enters through the 1-inch east port, as shown in fig.\,\ref{Fig1_Setup}. To increase the average number of reflections and thus the sensitivity, the aperture of this port was reduced to an internal diameter of 5 mm with a port reducer made of Spectralon\textregistered. We placed an optical trap, 8 mm thick, made of polyoxymethylene (POM), between the port reducer and the optical flat, which has internal V-grooves to reduce internal reflections. The volume between the west port reducer and the surface of the quartz window has a purge line connected to the west port adaptor to remove any air bubbles trapped there when the sphere is filled with liquid. This sphere is also equipped with an internal baffle that blocks the first bounce of light from reaching the photomultiplier.

To measure the reflected beam, we use the PMT R762P from Hammamatsu\textregistered~, mounted in the vertical position on the top of the north port as illustrated in fig.\,\ref{Fig1_Setup}A. P762-Y001 and R762P have an external diameter of 20~mm, a synthetic silica glass window, and a bialkaline photocathode. The main PMT is mounted in an optical cage system from Thorlabs\textregistered~ attached to a port adapter made of aluminum, which guarantees the PMT is always installed in the same position. This port adapter has a weir above the port reducer to ensure that the PMT window is constantly immersed in the liquid when the sphere is full of liquid. The PMT sits on the top of one of two port reducers made with Spectralon\textregistered. The first port reducer has an internal diameter of 16~mm and the second of 12~mm, and both have a thickness of 6~mm. The use of two-port reducers allows us to examine and eliminate any systematic errors resulting from photocathode uniformity.

When measuring the incident flux (fig.\, \ref{Fig1_Setup}B), the cap of the east port is removed and replaced with an adaptor holding the main PMT, R762P, which is mounted in the port. The PMT window faces the beam directly at a 0$^{\circ}$ angle as shown on the right side of fig.\,\ref{Fig1_Setup}. However, since the light reflected in the PMT window or photocathode can be reflected back by the sphere, it can increase the value of the incident flux. To minimize this effect, we installed a 3-inch wide and \diameter~1-inch tube made of anodized aluminum between the west and east port. In this position, the PMT can be moved forward and backward, with the total distance between the internal surface of the port reducer of the west port and the PMT window ranging between 50~mm and 90~mm. Measuring the light flux at different distances allows for checking the collimation of the incident beam and identifying any systematic associated with the PMT repositioning. This measurement was performed for both air and liquid. All the components of the optical system are placed within a black chamber made of aluminum and stainless steel, with the inside painted with anti-reflective Paint from TS-optics.

\subsubsection{Acquisition system}

Both PMTs operate in photon counting mode to minimize the impact of PMT gain variations on the measurement results. The operating voltages for the main PMT and monitor PMT were determined following the procedure described in ref.\,\cite{Hammamatsu} on page 148, resulting in values of +1200V and +1400V, respectively. The signal from both PMTs is fed into similar electronics, first into a fast filter amplifier with a differentiator set to a decay time of 10~ns, then discriminated, producing a NIM digital signal 8~ns wide, which is fed into a digital counter. 

To ensure accurate measurement results, we evaluated the time resolution of the acquisition system to determine the probability of pile-up. To conduct this evaluation, we arranged the main PMT in the configuration depicted in Fig.,\ref{Fig1_Setup}B and incrementally increased the light output from the LEDs. We then measured the photon flux ratio between the main PMT and the monitor PMT. The main PMT observed an average flux 20 times larger than the monitor PMT, leading to photon pile-up occurring earlier for the main PMT. Our measurements indicated that this value remained constant up to fluxes of 2.5$\times$10$^5$ detected photons per second (phd/s) in the main PMT. From these measurements, we estimated a data acquisition dead time of 18 ns after the electronic processing. To ensure the accuracy of the light flux measurements, we corrected for pile-up using the procedure outlined in \cite{Hammamatsu} (p.\ 131). Our pile-up correction had a maximum of 0.6\% for fluxes of 3.5$\times$10$^5$ phd/s, which was the maximum incident flux observed in this experiment.

\subsection{Measuring the incident and reflected flux}

In order to eliminate effects from possible instability of the LED output, the relative flux was calculated as the ratio between the count rates of the main and the monitor PMTs:
\begin{equation}
\Phi = \frac{N-N_0}{M-M_0},
\label{eq:fluxdef}
\end{equation}
where $N$ and $M$ are the number of observed photons (phd) recorded within 1-minute intervals by the main and the monitor PMTs, respectively, while $N_0$ and $M_0$ are the corresponding dark counts recorded during the same interval with all LEDs turned off. The dark count rate observed was between 30 and 70~phd/s for the air measurements and 180~phd/s for the water measurements. Next, all the mentioned fluxes will be relative as defined by eq.\,\ref{eq:fluxdef}.

To ensure the stability of the experimental setup and eliminate possible sources of error, a typical measurement sequence comprises several steps. First, we measure the dark count rates of both PMTs, denoted as $N_0$ and $M_0$, respectively. Next, we measure the count rates $N$ and $M$ for each LED in turn. Finally, we perform another measurement of $N_0$ and $M_0$. To ensure the reliability of the results, we repeat the sequence of seven LED measurements once or twice to check for any temporal evolution in the observed flux. Such changes could be due to variations in the PMT gain, fluctuations in the LED output, or slight changes in the system's geometry (as discussed in sec.\,\ref{refgasliquid}). 

To measure the throughput of the sphere in air, $H_{\mathrm{air}}$, we follow a two-step process. First, we measure the incident flux using the setup shown in Fig.,\ref{Fig1_Setup}B. Next, we mount the PMT in the north port and close the east port with a cap, ensuring that the sphere remains in the same position throughout the measurement.

The measurements in the liquid interface are performed right after the air measurements to ensure that the system's geometry and the sphere's reflectivity do not change significantly. After the measurement of $\Phi_R^{\mathrm{air}}$, the main PMT is removed without moving the sphere, and the liquid is poured into the sphere using a pipette. The PMT is slowly lowered until the PMT window is at the face of the internal surface of the sphere, and the weir of the east port is filled with liquid. The measurements were taken in sequence with the sphere filled at $\nicefrac{1}{3}$ (110\,m$\ell$), $\nicefrac{2}{3}$ (220\,m$\ell$), and full. 

Controlling the presence of bubbles in the sphere is crucial in this experiment, particularly near the ports. As such, after the liquid measurements to check the presence of air bubbles between the surface of the PMT and the liquid interface, we removed the PMT and placed it again in the same position using the same procedure, and the sequence of LED measurements was repeated. This method could identify bubbles in contact with the PMT as a significant shift in the observed flux. Additionally, we visually inspected the east port to exclude any bubbles present by removing the full sphere.

\subsection{The Photocathode reflectance}\label{sec:photocat_refl}

For both  $\Phi_I$ and $\Phi_R$ measurements, the light can be reflected in the PMT quartz window, photocathode, or the PMT internals \cite{MOTTA2005217}. While this light is mainly absorbed in the $\Phi_I$ measurement, it can be reflected in the sphere in the $\Phi_R$ measurement and still be detected, artificially increasing the measured value of the sphere reflectance. To account for this, we must consider the reflectance of the PMT. The probability of reflection in the PMT quartz window can be estimated accurately for both air and liquid interfaces since the refractive index of quartz is well-known for all the wavelengths used. However, the is unknown for the photocathode and PMT internals ($R_{ph}$) due to manufacturing process details that are unavailable. Therefore, we measured $R_{ph}$ directly using a dedicated setup. In the method used here, the throughput of the sphere is measured with the PMT mounted in the east port and compared with the throughput of an aluminum mirror with a known reflectance and mounted in the same position as the PMT. These results are further analyzed in a Monte Carlo simulation described in sec.\,\ref{sec:Simulations} to obtain the PMT reflectance.

To measure the photocathode reflectance, we mount the main PMT in the east port according to fig.\,\ref{Fig1_Setup}C, and the monitor PMT is mounted in the north port during the sequence of these measurements to measure the reflected fluxes. The main PMT now faces the incident beam with an angle of 8 degrees to prevent the light reflected in the PMT from escaping through the west port. Since the diameter of the PMT is smaller when compared with the diameter of the east port, we added a ring of Spectralon\textregistered~ reflector in front of the PMT (indicated in fig.\,\ref{Fig1_Setup}C) to increase the light output of the sphere. This reflector has an external diameter of 1~inch and an internal diameter of 12~mm. Next, the light flux from the PMT in the north port ($\Phi_R^{ph}$) is acquired for each LED using a similar data analysis process described previously, excluding the reference PMT. To calibrate the setup, we replaced the PMT mounted in the east port with a UV-Enhanced aluminum mirror PFSQ05-03-F01 with a known reflectance from Thorlabs\textregistered, keeping the same geometry with the same Spectralon\textregistered~ reflector ring in front of the mirror (fig.\,\ref{Fig1_Setup}D) and we measured the reflected flux in the north port ($\Phi_R^{A\ell}$).The reflectance of the aluminum mirror at an angle of incidence of 12$^{\circ}$ was provided by Thorlabs\textregistered, ranging from 87\% at 255 nm to 90.3\% at 490 nm. By comparing the measured $\Phi_R^{ph}$ and $\Phi_R^{A\ell}$ values, we could determine the reflectance of the PMT photocathode ($R_{ph}$). 

Each measured flux is divided by the incident flux $\Phi_I$ to obtain the throughput of the sphere in air, $H_{\mathrm{air}}$, for the photocathode reflectance, $H_{\mathrm{ph}}$, and the aluminum mirror reflectance, $H_{\mathrm{A\ell}}$. We compare these results with the simulated values, which are obtained using the method described in the next section, for each configuration shown in Figures \ref{Fig1_Setup}A, \ref{Fig1_Setup}C, and \ref{Fig1_Setup}D.
}\label{sec:Setup}

\section{The Equation of the Sphere and the Monte Carlo Simulations}{\subsection{The Equation of the Sphere}\label{subsec:EqOfTheSphere}

The throughput of an integrating sphere can be predicted using appropriate equations. For example, when the sphere wall is directly irradiated, the throughput of a sphere with two ports, an entrance, and an observing port, is given by (see eq.~8 in ref.\,\cite{Goebel:67}):
\begin{equation}
H = \frac{\eta_v R_{1} \left(1-R_v\right) }{1- R \left(1- \eta_v - \eta_e - \eta_a \right) - R_e \eta_e - R_v \eta_v},
\label{eq:SphereEquation}
\end{equation}
 where $R_1$ and $R$ correspond to the hemispherical reflectances of the sphere in the first and the subsequent internal reflections. The port fractions of the entrance (west) and viewing (north) ports are denoted by $\eta_e$ and $\eta_v$, respectively, and $\eta_a$ accounts for the losses due to light absorption. $R_e$ and $R_v$ correspond to the average reflectivity of the entrance and observing ports, respectively. Note that the equation presented here has an additional factor $\left(1-R_v\right)$ compared to the equation from ref.,\cite{Prokhorov:03} to account for the reflectance of the viewing port, which reduces its effective area by that amount. The calculation method for each component in this equation is described in appx.,\ref{AppendixD}.  

\subsection{The Monte-Carlo Simulation\label{MonteCarloSims}}

Although the equation for the sphere provided above is widely used to predict the response of integrating spheres, it has several limitations. Firstly, it does not account for partial fills with liquid or the roughness of internal surfaces, nor does it include Rayleigh scattering. Additionally, this equation assumes that all surfaces reflect light diffusely according to the Lambertian reflection model, which may not always be accurate. To address these limitations, we employed a Monte Carlo simulation to model light transport through the sphere. This simulation allows us to incorporate these additional details and investigate their impact on the sphere's output. Our simulation was conducted using ANTS2, a simulation and data processing package that specializes in modeling the transport of optical photons in scintillation-based detectors. By comparing the simulation results with the theoretical predictions from the sphere equation, we can better understand the factors that influence the sphere's performance.

In the Monte Carlo simulation, the internal volume of the integrating sphere is divided into three equally-sized regions to simulate the sequence of measurements at different fill levels: $\nicefrac{1}{3}$ capacity, $\nicefrac{2}{3}$ capacity, and full. Each region can be filled with a different material to accurately model the effects of different filling levels. 

In order to ensure accurate simulation results, it is crucial to provide a detailed description of the optical properties of all materials involved. This includes the optical properties of the Spectralon\textregistered\, the PMT's window and photocathode, and the liquid inside the sphere. The refractive indexes of the water and the fused silica glass (or fused quartz) are obtained using the Sellmeier dispersion formul\ae\, with the coefficients for the water measured by M.~Daimon and A.~Masumura \cite{WaterN} (at a temperature of 19$^{\circ}$C),  and by I.\ Malitson \cite{QuartzN} for fused silica glass. 

The refractive index of the Spectralon\textregistered\, was measured by Labsphere\textregistered\, to be 1.35 \cite{labsphereRI} (ASTM D-542), but information on its wavelength dependence is not available. We do have information on the dependence of the refractive index on the wavelength for the Teflon AF\textregistered~ \cite{10.1117/1.2965541}, a material similar to PTFE. Their results show that the difference in the refractive index between $\lambda=$490~nm and $\lambda=$255~nm is less than 0.02. Other similar materials also show similar differences in the refractive index at $\lambda=$490~nm and $\lambda=$255~nm (see ref.\,\cite{5411657}). Given this information, for the ultra-violet LEDs, we assumed a range of the refractive index between 1.35 and 1.37 in this analysis.

The reflectance model used in our simulations takes into account both specular and diffuse components. Since Spectralon\textregistered\ surfaces are known to be rough, we used the reflectance model proposed in our previous work \cite{SILVA201059}, which adapts equation \ref{BRIDF_mymodel} to account for roughness. To characterize the roughness of the Spectralon\textregistered\ surface, we employed a gonioreflectometer as described in ref.,\cite{ReflectanceJAPArticle}, and found that the surface profile follows the Trowbridge-Reitz distribution \cite{Reitz_1975} with a roughness parameter $\sigma_{\alpha}$ of 0.17. For all other materials, we assumed perfectly smooth surfaces with only a specular component.

Light absorption in the water might impact the results since, as shown in the eq.\,\ref{eq:SphereEquation}, it cannot be distinguished directly from a reduction in the reflectance of the sphere. A brief discussion of the most important water absorption measurements for the relevant wavelength ranges is described in appx.\,\ref{AppendixC}.  Although the absorbance of the pure water used in this experiment is not well-known, its very low conductivity of 0.163,$\upmu$S/cm, which is smaller than the conductivity of water used in previous studies by Irvin and Quickenden (0.430,$\upmu$S/cm) \cite{Absorption_Quickenden_and_Irvin} and Buiteveld (1.5,$\upmu$S/cm) \cite{Absorption_Buiteveld}, and comparable to the pure water from the Mason and Fry measurements (0.055,$\upmu$S/cm) \cite{Absorption_Mason16}, indicates high purity. We assumed a range for the absorption coefficient, $a_{\mathrm{abs, min}}$ to $a_{\mathrm{abs, max}}$, for each LED and considered the differences in the final results as a systematic effect. The higher values were taken from the works of Irvin and Quickenden ($\lambda<300$~nm) and Buiteveld ($\lambda>300$~nm), and the lower values were taken from the work of Mason and Fry (see Table \ref{LEDTable}).

The Rayleigh scattering in water has a negligible impact on the measurements. Our simulations indicate that the output of the sphere is not significantly altered unless the Rayleigh scattering coefficient, $\sigma_{\mathrm{ray}}$, is greater than 0.2 cm$^{-1}$. Calculations by Kröckel and Schmidt \cite{Krockel:14} predict a value of $\sigma_{\mathrm{ray}}<0.01$ cm$^{-1}$ for all the LED wavelengths used in this study, which is well below the sensitivity of the sphere. 

\subsection{Monte-Carlo results and comparison with the sphere equation}

To validate our Monte Carlo model, we conducted simulations of light propagation in the sphere under conditions where eq.\,\ref{eq:SphereEquation} is applicable, namely without a baffle and for a perfectly smooth wall surface. In these comparisons, we used a refractive index of 1.35 for the Spectralon\textregistered and assumed the refractive indices of water and fused silica to be those at $\lambda$=255\, nm (see tab.\,\ref{LEDTable}). We also used the PMT reflectance values obtained from simulations and measurements described in the following sections. The relative difference between the predictions made by the eq.\, \ref{eq:SphereEquation}, $H^{\mathrm{TIS\,eq}}$, and the simulation output, $H^{\mathrm{sim}}$,  is less than 4\% ($\left|H^{\mathrm{TIS\,eq}}-H^{\mathrm{sim}}\right|/H^{\mathrm{sim}}$) for $\rho$>0.6 and for both the liquid and air. 
Below 0.6, the sphere equation underestimates the sphere's output by 5\% at $\rho$=0.5 and 10\% at $\rho$=0. 

We next investigated the effect of surface roughness on the sphere's output, taking into account the roughness parameter $\sigma_\alpha$. Surface roughness can impact the sphere's output in two ways: (a) by increasing the shadowing effect and multiple scattering across the surface, which reduces the overall output, and (b) by reducing the probability of light being reflected back to the entrance port after the initial reflections inside the sphere. Our simulations showed that the effect of surface roughness is small for high albedos because (a) and (b) mostly cancel each other out, but for low albedos, surface roughness leads to a small increase in the sphere's output due to effect (b). For instance, in air, the sphere's output for $\rho=0.95$ and $\sigma_\alpha=0.17$ was largely unaffected by surface roughness, while in liquid, it decreased only by 0.8%.

Adding a baffle to the sphere has a more significant effect than roughness. For example, for $\rho=0.95$, we observed that introducing a baffle reduces by 4.2\% compared to a sphere without a baffle. To account for the effect of the baffle and the roughness in the eq.\,\ref{eq:SphereEquation}, we replaced the value of $R$ by $R^{b}$, where $b$ is an empirical constant, being $b$=1.07 for air and $b$ = 1.12 in the liquid.

\begin{figure}
 \begin{center}
     \includegraphics[width=8.4cm]{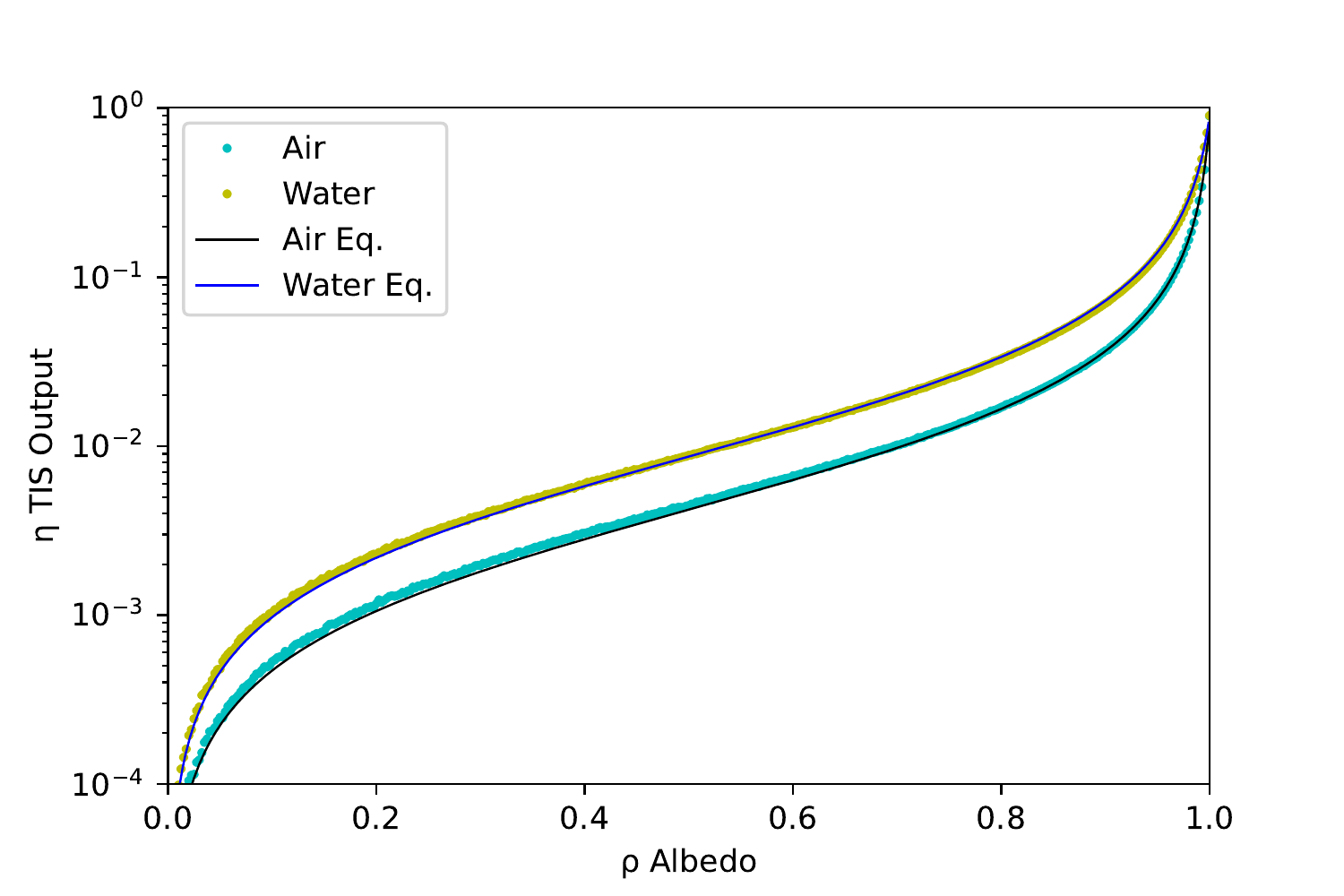}	
\caption{Comparison between the simulation of the sphere (data points) and the results using equation \ref{eq:SphereEquation} (full lines): the surface is assumed to be perfectly smooth, the baffle was removed and no light absorption in the water was considered. The refractive index of the Spectralon\textregistered\,is 1.35 and the quartz is 1.5048.}
  \label{ArtFig_Sim}
 \end{center}%
\end{figure}

When the sphere is full of water, the light absorption has the most considerable effect, especially for large values of $\rho$. For example, assuming the minimum absorption length for the LED of 255\,nm, we observed a reduction in the throughput of 7.3\% for $\rho=0.95$. To incorporate this effect into the sphere equation, we introduced a factor $f_a$ in eq.\,\ref{eq:SphereEquation}, which can be estimated using eq.,\ref{eq:TISabs}. 

After accounting for the effect of the baffle, roughness, and the absorption of water, the relative difference, defined as $\left|H^{\mathrm{TIS\,eq}}-H^{\mathrm{sim}}\right|/H^{\mathrm{sim}}$, is less than 2\% for the relevant range of measurements ($0.8<\rho<0.99$).

\subsection{Simulation of the PMT reflectance}

The Monte-Carlo method described before was applied to the measurement of the photocathode reflectance (fig.\,\ref{Fig1_Setup}C and \ref{Fig1_Setup}D). 
In the simulation results, we observed that the throughput with the PMT mounted in the east port ($H_{\mathrm{ph}}$) and then replaced by the aluminum mirror also mounted in the east port ($H_{\mathrm{A\ell}}$) is almost independent of the internal reflectance of the sphere $\rho$. The difference in the ratio ($H_{\mathrm{ph}}/H_{\mathrm{A\ell}}$) is less than 3\% between $\rho$=0.9 and $\rho$=0.99. Considering both measurements simultaneously, we can measure $R_{ph}$, almost independently on the albedo $\rho$ of the sphere.}\label{sec:Simulations}

\section{Results}{\subsection{Throughput of the sphere in air and liquid}\label{refgasliquid}

\begin{figure}[htpb]
 \begin{center}
     \includegraphics[width=8.4cm]{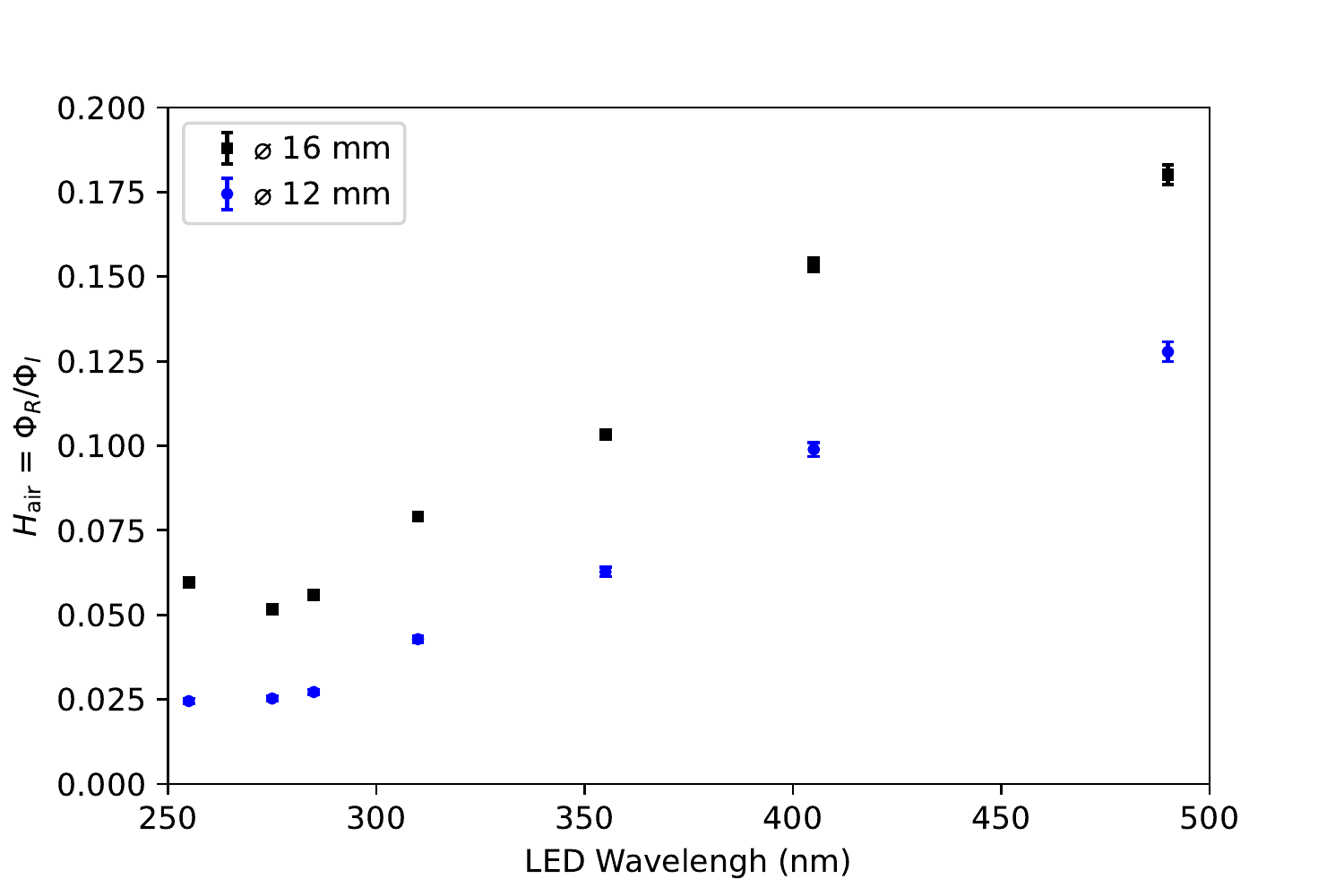}	
\caption{Throughput of the sphere in the air,  $H_{\mathrm{air}}$: the ratio between the reflected flux $\Phi_R^{\mathrm{air}}$ and the incident flux $\Phi_I^{\mathrm{air}}$ is shown as a function of the LED wavelength for the two port reducers used in the north port. }
  \label{ArtFig4_IncidentBeamFinalResultsFact1}
 \end{center}%
\end{figure}

The observed throughput of the sphere corresponds to the ratio $\Phi_R/\Phi_I$. However, the measured incident flux, $\Phi_I^{\mathrm{mea}}$, is slightly less than $\Phi_I$, due to the reflection of the incident light from the PMT window and photocathode. 

To obtain the true throughput of the sphere, we need to correct for the fact that the measured incident flux, $\Phi_I^{\mathrm{mea}}$, is slightly lower than the actual incident flux $\Phi_I$ due to the reflection of the incident light from the PMT window and photocathode. Taking into account that this reflected light goes back to the entrance window and can be reflected again in the direction of the PMT, one can write:
\begin{equation}
\Phi_I^{\mathrm{mea}} = \frac{1-R_{v}}{1-R_{v}R_{e}} \Phi_I,
\label{eq:incidentfluxcorrection}
\end{equation}
where $R_{v}$ and $R_{e}$ are the probabilities of reflection at the PMT and the entrance port, respectively. 
Consequently, the true throughput, $H$, is obtained by multiplying $\Phi_R/\Phi_I^{\mathrm{mea}}$ by the correction factor $\frac{1-R_{v}R_{e}}{1-R_{v}}$. The expressions for both $R_{v}$ and $R_{e}$ are provided in appx.\,\ref{AppendixE}.

 The throughput $H_{\mathrm{air}}$ is presented in fig.\,\ref{ArtFig4_IncidentBeamFinalResultsFact1} as a function of the LED wavelength for the two diameters of the north port reducer. Consistent with expectations, the throughput decreases at shorter wavelengths, indicating a lower reflectance of the sphere. 

The uncertainty of $H$ in air, $\sigma_{H_{\mathrm{air}}}$, is estimated by propagating the uncertainties in the measurements of both fluxes, which is dominated by systematic uncertainties since, for most of the LEDs, the statistical uncertainties associated with Poisson fluctuations are less than 0.1\%. The primary source of uncertainty affecting $\Phi_R^{\mathrm{air}}$ and $\Phi_I^{\mathrm{air}}$  is the temporal drift in the response of the PMTs and LEDs. To estimate the uncertainty $\sigma_{H_{\mathrm{liq}}}$, we repeated the sequence of the measurements three times with a 15-minute difference delay between each measurement, as described in sec.\,\ref{sec:Setup}. The relative uncertainty was similar across all LEDs, so we took the average standard deviation across all LEDs. We found that the uncertainty in $\Phi_R^{\mathrm{air}}$ was 0.6\%, while that in $\Phi_I^{\mathrm{air}}$ was 0.9\%.

\begin{figure}
 \begin{center}
     \includegraphics[width=8.4cm]{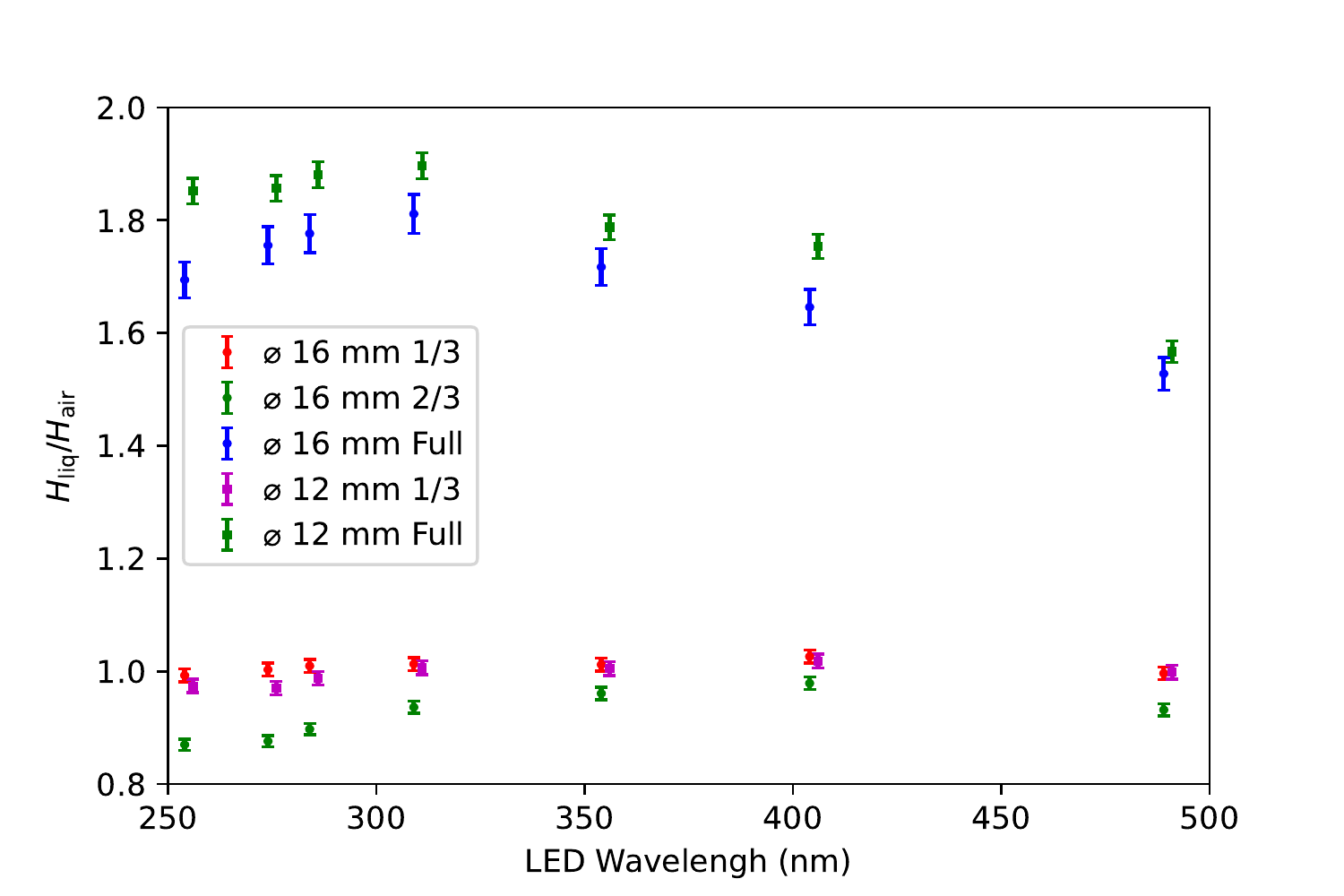}	
\caption{The ratio between the throughput in the liquid $H_{\mathrm{liq}}$ and in the air $H_R^{\mathrm{air}}$ for the full sphere and the sphere filled at $\nicefrac{1}{3}$ and $\nicefrac{2}{3}$ of its volume.}
  \label{Art5_ResultsWater}
 \end{center}%
\end{figure}

The measurement of $H_{\mathrm{liq}}$ differs from that in air because the sphere must be emptied and dried to remove any liquid present in the purge lines of the west port between the measurement of $\Phi_R^{\mathrm{liq}}$ and $\Phi_I^{\mathrm{liq}}$. This process adds new uncertainties, which are difficult to estimate. To minimize these uncertainties, we adopted a method in which we first obtained the ratio $\Phi_R^{\mathrm{liq}}/\Phi_I^{\mathrm{air}}$ and then correct it using the factor $\mathcal{M}_\mathrm{inc}$ to account for the different probability of the light to be refracted through the entrance window into the sphere. As such, $H_{\mathrm{liq}}$ is given by:
\begin{equation}
    H_{\mathrm{liq}} = \frac{\Phi_R^{\mathrm{liq}}}{\Phi_I^{\mathrm{liq}}} = 
    \left(\frac{\Phi_R^{\mathrm{liq}}}{\Phi_I^{\mathrm{air}}}\right) \mathcal{M}_\mathrm{inc}\left(n_{\mathrm{SiO_2}}, n_{\mathrm{liq}}\right),
    \label{liqmeasurement}
\end{equation}
where reflected flux in liquid, $\Phi_R^{\mathrm{liq}}$, represents the sphere filled at $\nicefrac{1}{3}$, $\nicefrac{2}{3}$, or to the top. The ratio $\mathcal{M}_\mathrm{inc}\left(n_{\mathrm{SiO_2}}, n_{\mathrm{liq}}\right)$ corresponds to the ratio between the incident flux in air and in the liquid. It was calculated considering the refractive indexes of the entrance window and the liquid. It is 1 when the sphere is filled at \nicefrac{1}{3}, and it ranges from 0.968 ($\lambda=$255\,nm) to 0.962 ($\lambda=$490\,nm) when the sphere is at \nicefrac{2}{3} capacity and full.  To account for possible systematic uncertainties, we also measured $\mathcal{M}{\mathrm{inc}}$ experimentally by directly measuring the incident flux in the liquid, $\Phi_I^{\mathrm{liq}}$, right after the measurement of the incident beam $\Phi_I^{\mathrm{gas}}$ with the PMT reflectance corrections applied (see eq.\,\ref{eq:incidentfluxcorrection}). The average difference between the observed and calculated value of the ratio, $\left({\Phi_I^{\mathrm{air}}}/{\Phi_I^{\mathrm{liq}}}\right)$, was (-0.18$\pm$0.22). As such, since the calculated value is not affected by experimental uncertainties, we used the calculated value of $\mathcal{M}{\mathrm{inc}}$ in our analysis.

Figure \ref{Art5_ResultsWater} shows the ratio $H_{\mathrm{liq}}/H_{\mathrm{air}}$ for all the LEDs.  For a full sphere, the ratio is consistently larger than 1.5 across all wavelengths. However, for partial fills, the ratio drops below 1 due to additional reflections at the interface between the liquid and air, which increases the effective path length of light in the sphere. These reflections lead to a reduction in the throughput of light, which is reflected in the smaller values of the ratio for partial fills.

To ensure the reliability of our measurements in liquid, we repeated the sequence of LED measurements three times for each volume of liquid, looking for any possible temporal changes in the observed fluxes. The liquid measurements have two additional sources of systematic uncertainties that might cause such dependence: a) the dissolution of impurities present in the sphere into the liquid (see appx.\,\ref{AppendixB} for a discussion on the cleaning procedure), and b) the presence or formation of air bubbles in the east port or the north port. Their impact on the measurements is limited since the throughput decreased by only an average of 0.6\% between the first and last measurement, despite a 30-minute time difference.

\subsection{Reflectance of the photocathode}

\begin{figure}
 \begin{center}
     \includegraphics[width=8.4cm]{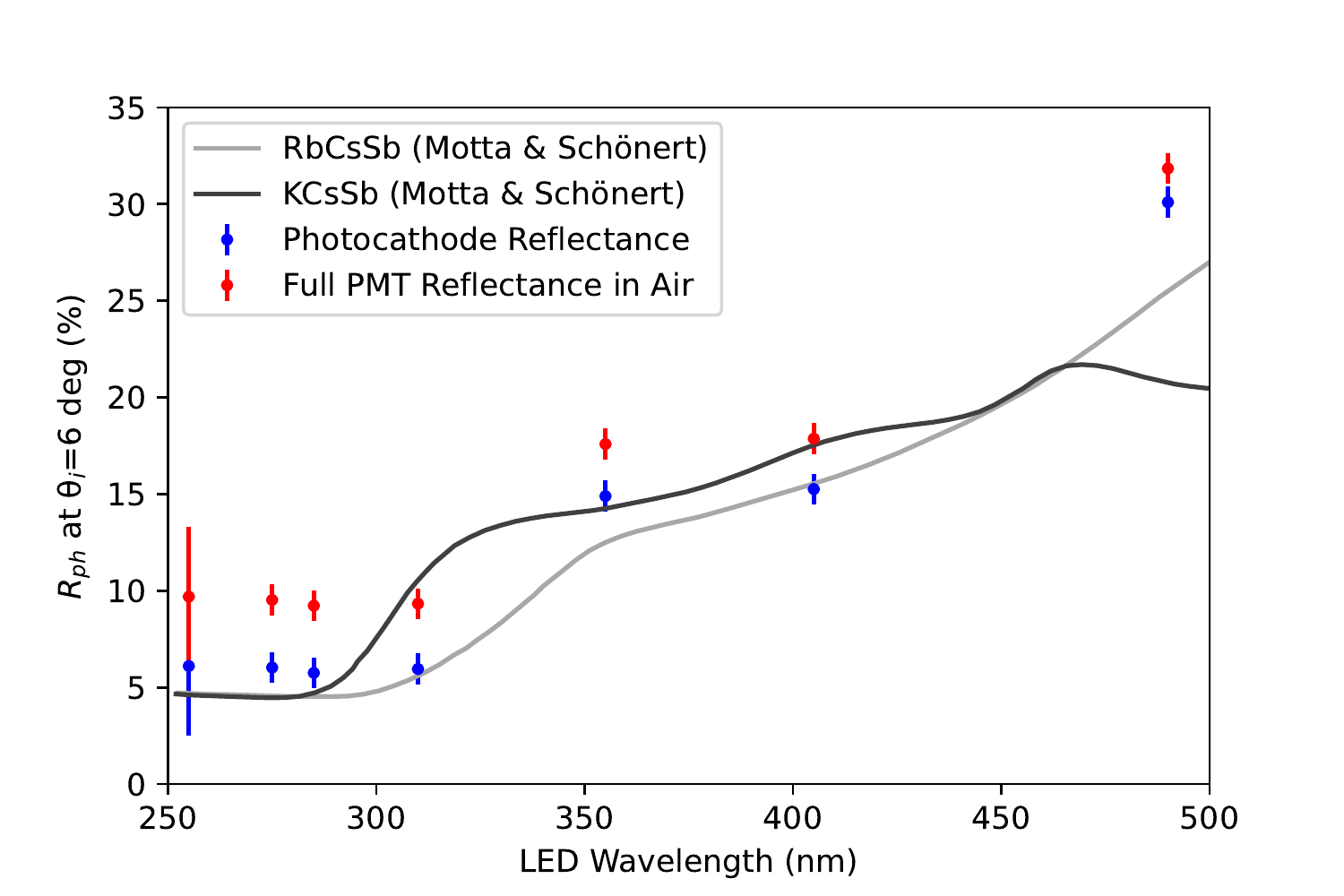}	
\caption{Measurement of the photocathode reflectance: (blue) observed photo-cathode reflectance as a function of the wavelength; (red) contribution from the PMT quartz window included; (gray and black curves) observations from Motta \& Schönert of the dependence of the PMT reflectance with the wavelength \cite{MOTTA2005217}. The error bars are five times larger for easier visualization.}
  \label{PhotocathodeMeasurements}
 \end{center}%
\end{figure}

The reflectance of the photocathode, $R_{ph}$, is a necessary input in both the equation of the sphere (eq.\,\ref{eq:SphereEquation}) and in eq.\,\ref{eq:incidentfluxcorrection} for correcting the incident flux. The reflectance of the photocathode, $R_{ph}$ for each LED is obtained using the set of observed quantities $\mathcal{H} = [H_\mathrm{air}, H_\mathrm{ph}, H_{\mathrm{A}\ell}]$ mentioned in sec.\,\ref{sec:photocat_refl} with $H_\mathrm{air}$ being an independent measurement from the one mentioned in sec.\,\ref{refgasliquid}. These three quantities are compared in a $\chi^2$ minimization with the results obtained with the simulations, $\mathcal{S}$, for the respective geometry in order to obtain both the photocathode reflectance, $R_{ph}$, and the multiple-scattering albedo of the sphere, $\rho$, for each LED wavelength:
\begin{equation}
\chi^2(\rho, R_{ph}) = \sum_{i} \frac{\left(\mathcal{S}_i-\mathcal{H}_i\right)^2}{\sigma_i^2},
\end{equation}
where $i$ runs over the three measurements of the throughput, and $\sigma_i$ is the estimated uncertainty for each measurement. For $\sigma_i$, we made the same assumptions made in sec.\,\ref{refgasliquid}.

The result obtained in this minimization for the reflectance of the photocathode, $R_{ph}$,  is shown in fig.\,\ref{PhotocathodeMeasurements}. Additionally, we show the total reflectance of the PMT in an air interface, which includes the reflection in the PMT window and the multiple possible reflections between the photocathode and the surface of the PMT. Finally, we compared these results with those obtained from Motta and Schönert \cite{MOTTA2005217} for two bi-alkaline PMTs with a glass window. As shown, the reflectance of the photocathode strongly depends on the wavelength of the light and compares well with the results obtained by the different authors. Moreover, the absolute difference in the albedo value, $\rho$, obtained with this method and the method described next is on average only 4$\times$10$^{-3}$. Therefore, the experimental values obtained here are next used in the data analysis.

\subsection{Multiple-scattering albedo}

To obtain $\rho$, its value used in the Monte Carlo simulations (section \ref{sec:Simulations}\ref{MonteCarloSims}) is adjusted to obtain the best match between $H^{\mathrm{sim}}$ and $H^{\mathrm{obs}}$. For each measurement, we adjust $\rho$ in such a way that $\vert H^{\mathrm{sim}} - H^{\mathrm{obs}} \vert <$5$\times$10$^{-6}$. For the partial fills ($\nicefrac{1}{3}$ and $\nicefrac{2}{3}$ volumes), $\rho$ can have different values in the liquid and gaseous phases, $\rho_{\mathrm{liq}}$ and $\rho_{\mathrm{air}}$. Since $\rho_{\mathrm{air}}$ is less affected by systematic uncertainties, we adjusted $\rho_{\mathrm{liq}}$ for these geometries while fixing $\rho_{\mathrm{air}}$ to the value obtained with the empty sphere.

The uncertainty in $\rho$ was obtained by error propagation of the sphere equation (see eq.\,\ref{eq:SphereEquation}). The contributions assumed are from $\sigma_H$, calculated in the previous section, the uncertainty in the refractive index of the PTFE, $\sigma_{n_{\mathrm{PTFE}}}$, the uncertainty in the absorption length $\sigma_{a_{\mathrm{abs}}}$, and the uncertainty in the reflectance of the photocathode. To determine $\sigma_{n_{\mathrm{PTFE}}}$, we considered the range between 1.35 and 1.37 for the UV LEDs ($\lambda$<400\,nm). For the absorption length in water, $a_{\mathrm{abs}}\pm \sigma_{a_{\mathrm{abs}}}$ is given by the range of absorption lengths ($a_{\mathrm{abs, min}}$ to $a_{\mathrm{abs, max}}$) shown in tab.\,\ref{LEDTable}. 

The results for $\rho$  are shown in fig.\,\ref{ArtFig6_ResultsFit} for the two port reducers in both liquid and air interfaces. For the full sphere, the average difference between the albedo in air and water,  $<|\rho_{\mathrm{air}} - \rho_{\mathrm{liq}}|>$, is 0.9$\times 10^{-3}$ with the maximum difference, $\mathrm{max}(|\rho_{\mathrm{air}} - \rho_{\mathrm{liq}}|)$, being 2.5$\times 10^{-3}$. These values increase to 1$\times 10^{-3}$ and 4$\times 10^{-3}$ for the $\nicefrac{1}{3}$ volume. 

We do not show the results for the volume $\nicefrac{2}{3}$ because these measurements were affected by air bubbles formed in the east port. These bubbles could not be appropriately removed because the liquid level was below the level of the purging lines of the east port. The presence of these bubbles reduces the observed flux by almost the same fraction across all the LEDs.

\begin{figure}
 \begin{center}
     \includegraphics[width=8.4cm]{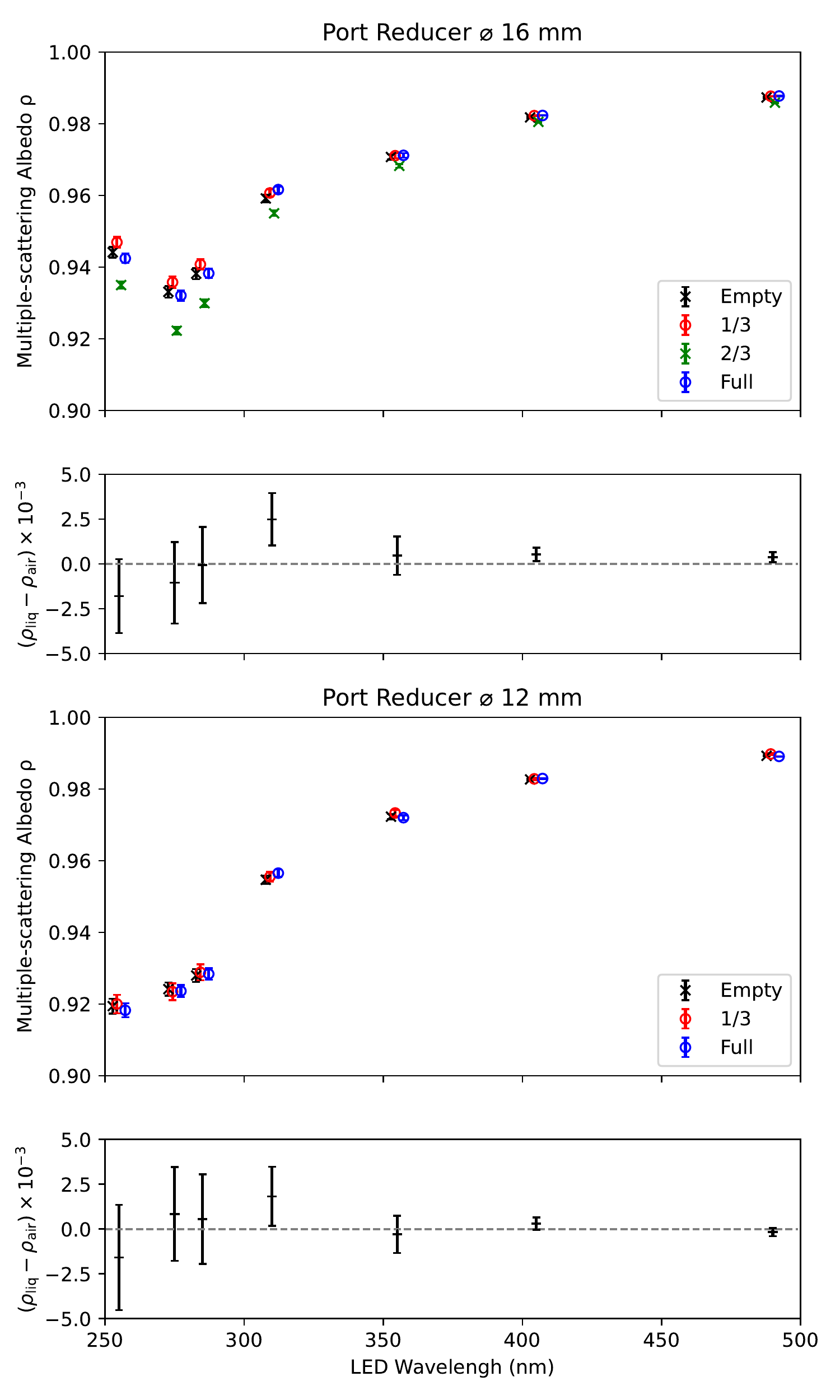}	
\caption{Results for the reflectance with the internal Lambertian model: the multiple-scattering albedo dependence with the LED wavelength is shown for the different geometries and the two port reducers. The different geometries have a small shift (<2~nm) in the horizontal axis to improve visualization.}
 \label{ArtFig6_ResultsFit}
\end{center}%
\end{figure}

\subsection{Bi-hemispherical reflectance}

\begin{figure}
 \begin{center}
     \includegraphics[width=8.4cm]{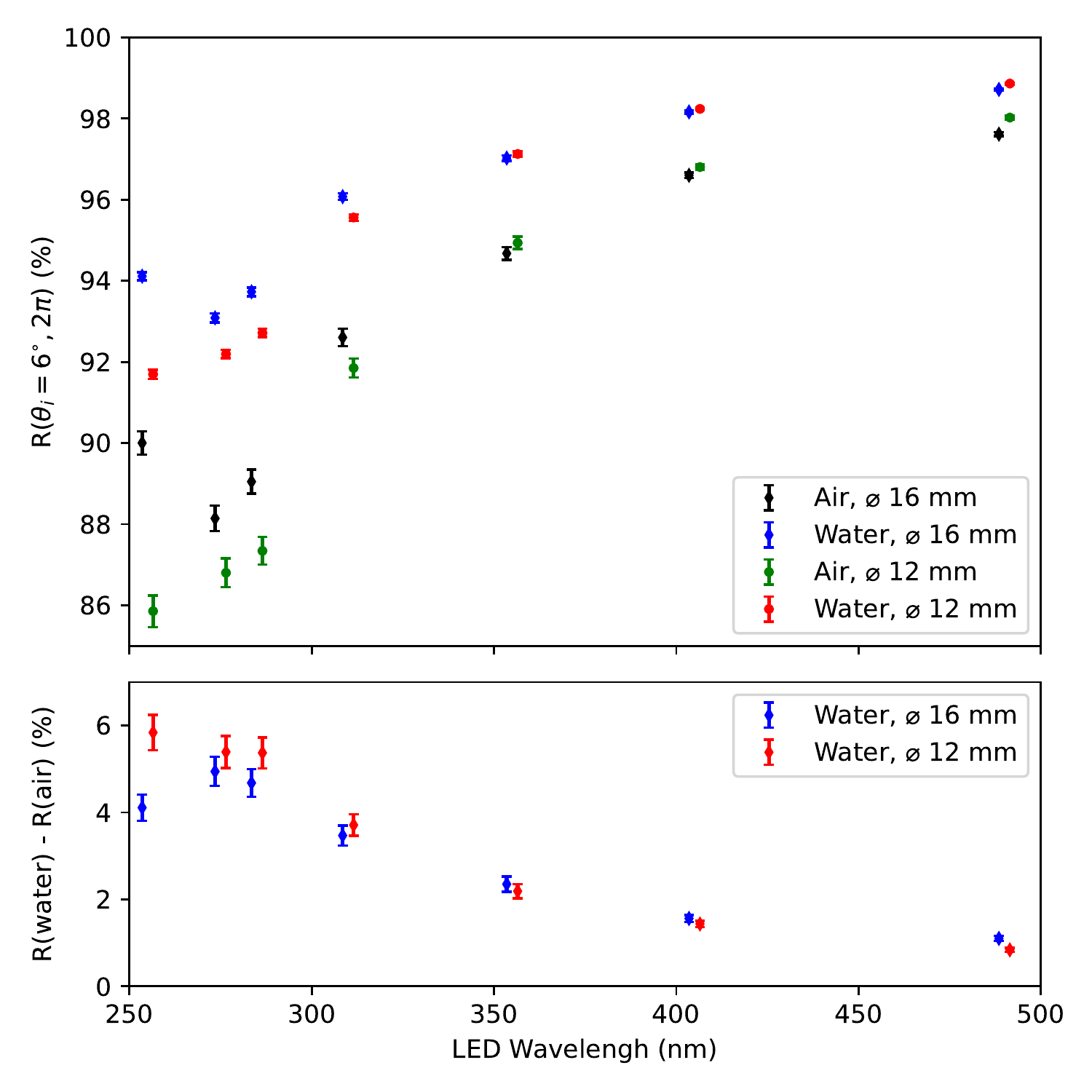}	
\caption{Dependence of the with the wavelength of directional-hemispherical reflectance calculated with the ANTS simulation $R\left(\theta_i=6^{\circ};2\pi\right)$. }
  \label{ArtFig8_DHR}
 \end{center}%
\end{figure}

The hemispherical reflectance of a diffuser can be characterized with a 6$^{\circ}$ directional-hemispherical factor $R\left(\theta_i, \phi_i;2\pi\right)$ which can be directly measured using an integrating sphere as exemplified in the Weidner and Hsia work \cite{Weidner:81}. $R\left(\theta_i, \phi_i;2\pi\right)$ is defined as the ratio between the reflected flux measured over the whole hemisphere above the surface and the incident flux assuming a direction of incidence defined by the angles ($\theta_i$, $\phi_i$) \cite{nicodemus, Judd:67}:
\begin{equation}
R\left(6^{\circ}, 0;2\pi\right) = \frac{1}{\pi} \int_{2\pi} f_r \left(\theta_i = 6^{\circ}, \phi_i = 0;\theta_r, \phi_r \right) \mathrm{d}\Omega_r,
\label{eq:DR}
\end{equation}
where $\mathrm{d}\Omega_r$ is given by equation \ref{eq:dOmega}. Assuming that the surface is smooth, this integral is given by:
\begin{equation}
\begin{split}
R\left(6^{\circ}, 0;2\pi\right) =& F\left(6^{\circ}; \frac{n_2}{n_1}\right)  +  \frac{\rho}{1 - \rho \overline{\mathcal{F}}_{\nicefrac{n_1}{n_2}}}\times \\ & \times \left[1 - F\left(6^{\circ}; \frac{n_2}{n_1}\right) \right]\left[ 1 - \overline{\mathcal{F}}_{\nicefrac{n_1}{n_2}} \right].
\label{eq:hemisph}
\end{split}
\end{equation}

This is the well-known Saunderson correction (eq.\,6 in \cite{Saunderson:42} and eq.\,2 in ref.\,\cite{Blevin:62}) with the factor $k_1=F\left(6^{\circ}; \nicefrac{n_2}{n_1}\right)$ and $k_2=\overline{\mathcal{F}}_{\nicefrac{n_1}{n_2}}$.

Fig.\,\ref{ArtFig8_DHR} presents the  results for $R\left(\theta_i=6^{\circ};2\pi\right)$ in a liquid and gaseous interfaces. To account for surface roughness, $R$ is obtained using the Monte Carlo simulation based on ANTS (see sec.\,\ref{sec:Simulations}). As shown, the reflectance in liquid water is larger for all LEDs and geometries. The difference in $R$ between the liquid and air is between 6\% for $\lambda$<300~nm, and it decreases to 0.84\% for $\lambda$=490~nm. The diffuse reflectance causes this increase since the specular reflectance decreases in the liquid due to the smaller difference in the refractive index between the liquid and the PTFE/quartz. The results aligned with the observations from ref.\,\cite{Voss:06}, which showed an increase of 2\% using a He-Ne laser ($\lambda$=632.8\,nm) as a light source.

When the roughness parameter, $\sigma_{\alpha}$, is reduced from 0.17 to 0, the results from the Monte Carlo agree well with eq.\,\ref{eq:hemisph}. The average difference between the reflectance of the smooth surface  $\sigma_{\alpha}=0$ and a rough surface $\sigma_{\alpha}=0.17$ was -0.14\% for the air and +0.08\% for the liquid. These findings are consistent with the predictions made by ref.\,\cite{doi:10.1063/1.3318681}, which suggests that surface roughness has minimal effect on integrated reflectances, although it does affect the angular distribution of reflected light.}\label{sec:Results}

\section{Discussion}{The reflectance values in air obtained here are lower than the results reported by Weidner et al.~\cite{Weidner:81, Weidner:85}, which shows reflectance values above 97\% for $\lambda$>250\,nm. As reported by some authors \cite{Georgiev:07, Shaw:07}, the Spectralon reflectance degrades over time, even when kept under cleaning room conditions. This decrease, stronger for newer samples, is caused by the absorption of impurities, especially aromatic hydrocarbons. We observed this effect as the throughput of the sphere decreased roughly by 50\% for 255\,nm compared with the initial tests in early 2020. Nonetheless, this decrease in reflectance is advantageous for this work since it increases the range of multiple-scatter albedos that could be assessed with this setup.

The value of the multiple-scattering albedo, $\rho$, can be obtained analytically, assuming that the roughness has no significant impact on the value of the hemispherical reflectances, which, as discussed earlier, is a valid assumption even for moderately rough surfaces. If the value of the directional-hemispherical factor $R\left(\theta_i;2\pi\right)$ has been measured for a specific angle of incidence, the value of $\rho$ is obtained by replacing the value of $f_r$ defined in eq.\ \ref{BRIDF_mymodel} in the directional-hemispherical integral (eq.\,$\ref{eq:DR}$). Solving for $\rho$ results in:
\begin{equation}
\rho = \frac{R\left(\theta_i;2\pi\right)-F\left(\theta_i; n\right)}{1-F\left(\theta_i; n\right) - \overline{\mathcal{F}}_{\nicefrac{1}{n}} \bigl[ 1-R\left(\theta_i;2\pi\right) \bigl]},
\label{EqMultipleScatteringAlbedo}
\end{equation}
where $n=n_2/n_1$ corresponds to the relative refractive index.

We assumed that the multiple-scattering albedo value, $\rho$, does not depend on the angle at which the light refracts into the diffuser. However, Chandrasekhar predicted the dependence of the diffuse reflection with the angle of incidence in his work on Radiative Transfer work (\cite{Wolff:94, Chandrasekhar1960}). In semi-infinite diffusers, the diffuse properties depend on the single-scattering albedo, corresponding to the probability that the light is not absorbed during two consecutive scatters and on the angular distribution of scattered light. We implemented the Chandrasekhar model for the isotropic scattering and tested it with our data. This was achieved by replacing the Lambertian model with the eq.\,123 in chapter\,3 of ref.\,\cite{Chandrasekhar1960}. The performance of this model is similar to the model presented before, and since it is more complex and computationally more expensive, it has no clear advantage in predicting the sphere reflectance in the liquid compared to the Lambertian model. Nonetheless, the distinction between these two models is more substantial when the surface is illuminated or observed at a larger angle. An integrating sphere is not sensitive to this because the angle of incidence closely follows the Lambertian law disfavouring large angles of incidence and requiring a different geometry to assess the performance of these models at large angles of incidence.

The results from fig.\,\ref{ArtFig8_DHR} contradicts the common knowledge that surfaces look darker when wet. However, this occurs when a layer of liquid covers the diffuser adding a new optical interface where the light can be reflected back to the diffuser \cite{Lekner:88}. The results presented here also show this effect since the ratio ($\Phi_R^{\mathrm{liq}}/\Phi_R^{\mathrm{air}}$) was smaller than 1 when the sphere was filled at \nicefrac{1}{3} and \nicefrac{2}{3}. 

Another cause of the decrease in reflectance in a liquid interface is when the material is porous and can absorb the liquid, which alters the multiple-scatter albedo of the diffuser. As reported in ref.\,\cite{Twomey:86}, such absorption changes the refractive index of the diffuser's bulk bringing it closer to the refractive index of the liquid and, consequently, increasing the forward scattering. As such, the light penetrates farther, increasing the absorption probability. Spectralon\textregistered\ is a well-known porous material with a density between 1.25 and 1.5~g/cm$^3$, smaller when compared with crystalline PTFE (2.2~g/cm$^3$). However, in the case of measurements with water, this porosity is irrelevant, as it is also hydrophobic with a water permeability of only 0.001\% (ASTM D570 test made by Labsphere\textregistered\, \cite{labsphereRI}). However, when we performed these measurements with cyclohexane \ce{C6H12}, $\rho$ decreased as much as 10\% for 255~nm and 1\% for 490~nm. Since \ce{C6H12} is an apolar liquid, it soaks the Spectralon\textregistered\, entering the air voids in the material and changing its optical properties. We tested this hypothesis by soaking a 2~g piece of Spectralon\textregistered\, with \ce{C6H12} during 24~h, in which we observed a total mass increase of 12\%. Further studies are necessary to fully describe the reflectance when the liquid is absorbed by the diffuser.

\section{Conclusion}\label{sec:Conclusion}

This work has demonstrated that a single parameter, the multiple-scatter albedo $\rho$, can predict the diffuse reflectance in both air and liquid water. To show this, we build a set-up composed of a total integrating sphere capable of measuring the reflectance in both air and liquid. Then, we developped a detailed Monte-Carlo model to predict the sphere's throughput for a specific value of $\rho$, and we compared it with the sphere equation adapted to the measurement in liquid. Finally, the Monte-Carlo simulation was compared with the obtained data to get $\rho$ for a specific configuration, and we calculated the difference between the albedo in the air and water $\Delta\rho = \rho_{\mathrm{air}} - \rho_{\mathrm{liq}}$.
For the full sphere, the average difference between the albedo in air and water,  $<|\Delta\rho|>$, is 0.9$\times 10^{-3}$ with the maximum difference, $\mathrm{max}(|\Delta\rho|)$, being 2.5$\times 10^{-3}$. This difference, when normalized to the respective uncertainty ($<|\Delta\rho/\sigma_{\Delta\rho}|>$) was 0.7, indicating good agreement between the values of $\rho$ in liquid water and air. 

We also showed that the parameter $\rho$ can be obtained using eq.\,\ref{EqMultipleScatteringAlbedo} with the hemispherical reflectance of the surface measured in air at a specific angle of incidence. This allow us to predict the reflectance in a the liquid without without measuring it precisely for that interface, , thereby avoiding complications related to liquid purity, absorption, and the adaptation of the setup. Overall, these findings provide valuable insights for the design and optimization of particle detectors that rely on liquid interfaces.

}\label{sec:Discussion}

\appendix
\section{The Fresnel equations}\label{AppendixA}

The Fresnel equations set both the intensities of the reflected and refracted waves. Using these equations we can obtain the reflectivity $F$ given by (eq. 32 and 33 of ref.\,\cite{Born:382152}):
\begin{align}
\begin{split}
F\left(\theta_i;n, \alpha\right) = \frac{\tan^2\left(\theta_i - \theta_t\right)}{\tan^2\left(\theta_i + \theta_t\right)} \cos^{2}\alpha +  \frac{\sin^2\left(\theta_i - \theta_t\right)}{\sin^2\left(\theta_i + \theta_t\right)} \sin^{2}\alpha,
\label{Eq:Fresnel}
\end{split}
  \end{align}
where $\theta_t = \arcsin\left(\sin\theta_i /n \right)$ and $\alpha$ is the angle in which the electric field vector of the incident wave makes with the plane of incidence.

When the radiation propagates uniformly in all directions with random polarization (isotropic irradiation), the average reflectance is given by $\overline{\mathcal{F}}_n$ defined previously in the eq.\,\ref{IntegralFo}. Stern \cite{Stern:64} obtained the solution for this integral. For $n>1$, it is given by:
\begin{equation}
\begin{split}
\overline{\mathcal{F}}_n = & n^2 \bigg[ \frac{3n^2+2n+1}{3n^2 \left(n+1\right)^2} + \frac{n^2+2n-1}{\left(n^2+1\right)^2\left(n^2-1\right)}  \\
& + \frac{n^2 +1}{\left(n^2 - 1\right)^2} \log n - \frac{\left(n^2-1\right)^2}{\left(n^2 + 1 \right)^3} \log\frac{n^2+n}{n-1} \bigg].
\label{integralresultado}
\end{split}
\end{equation}
To get the values for $n<1$, we make use of the  relation \cite{Stern:64, Lekner:88}:
\begin{equation}
\overline{\mathcal{F}}_n = 1-n^2\left(1-\overline{\mathcal{F}}_n\right).
\label{relacaoinverso}
\end{equation}

Most dielectric materials have a refractive index smaller than 2.0. In that case, the integral \ref{integralresultado} can be approximated to:
\begin{equation}
\overline{\mathcal{F}}_n \simeq 1 - n^2 \left(a + \frac{1-a}{2b^{n-1}-1} \right), \quad 1<n<2,
\end{equation}
with  $a = 0.0364$ and $b = 3.280$  with the same relation (eq.  \ref{relacaoinverso}) for $n<1$. The relative error of this approximation is less 6$\times$10$^{-4}$ for $0.5<n<2.0$.

\section{Measurements with Water}\label{AppendixB}

In this section, we provide some details related to the water measurement.

The water used in this study is ThermoFisher spectroscopy-grade ACS water, identified by catalog number 43338 and lot number A0429271. The electrical conductivity of the liquid is measured to be 16.3 $\mu$S/m.

\subsubsection*{Cleaning procedure}

The sphere is thoroughly cleaned before each measurement using the following procedure: (i) nitrogen is sprayed over the surface to remove any particulates, (ii) the sphere is rinsed with ultrapure water, (iii) it is cleaned with propanol, (iv) it is heated in an oven at a temperature of 45~$^{\circ}$C for 2 hours, (v) vacuum is applied to the sphere at 1~mbar for 3 hours, and (vi) the sphere is finally bathed in an argon atmosphere for at least one hour.

\subsubsection*{Water absorption Coefficient}\label{AppendixC}

The reported value of water absorption coefficient in the range of 255-490~nm, as reported by different authors, can differ up to two orders of magnitude \cite{Absorption_Quickenden_and_Irvin, Absorption_Buiteveld, Absorption_Mason16}. This inconsistency is because water exhibits minimal absorption above 250 nm, so the measurements are prone to systematic uncertainties that are difficult to control. The differences have been attributed to factors such as varying water purity levels, differences in Rayleigh scattering predictions, and variations in measurement methodologies \cite{Fewell:19}. One of the most cited studies on the absorbance of liquid water is from Irvin and Quickenden for $\lambda$<320~nm \cite{Absorption_Quickenden_and_Irvin}. They used a differential path length method, in which the absorbance was measured using two cells with different sizes, and then the contribution from Rayleigh scattering was removed. Buiteveld \cite{Absorption_Buiteveld} made another significant measurement above 300 nm using a submersible absorption meter \cite{10.1117/12.190123} and reported a minimum absorption of $a$=5.3 km$^{-1}$ at 386~nm. However, these measurements are dependent on the considered value of Rayleigh length, which is not well established. In 2016, Mason and Fry \cite{Absorption_Mason16} measured the absorption coefficients above 250~nm with an integrating cavity, a new technique independent of the scattering effects \cite{Fry:92}. They observed a minimum absorbance of 0.81\,km$^{-1}$ at 344~nm.

\section{Equation of the sphere}\label{AppendixD}

As discussed in the section, \ref{sec:Simulations}\ref{subsec:EqOfTheSphere}, the output of the sphere can be predicted with:
\begin{equation}
H = \frac{\eta_v R_{1} \left(1-R_v\right) }{1- R^{b} \left(1- \eta_v - \eta_e - \eta_a \right) - R_v \eta_v}.
\label{TISeq_mod}
\end{equation}
Here, we describe how to obtain each quantity in this equation.

\subsection{The port fractions}

The port fraction, $\eta_i$, for each port $i=(e, v)$ can be obtained with:
\begin{equation}
 \eta_i = \frac{1}{2}\left[1 - \sqrt{1-2\frac{r_i^2}{r^2}}\right],
\end{equation}
where $r$ corresponds to the sphere's radius and $r_i$ to the radius of the entrance or viewing port. $\eta_a$ describes the effect of the water absorption and is discussed next.

\subsection{The reflectance of entrance port}

Photons can return to the entrance port (the west port) during the multiple reflections inside the sphere. Upon reaching the quartz window, these photons may undergo reflection or refraction. Based on Monte Carlo simulations (sec.\, \ref{sec:Simulations}), we estimated a maximum probability of 4\% for photons to return to the entrance port. However, due to the window's geometry, the photons' incident direction is nearly perpendicular to the window, resulting in refraction and exit from the sphere. Furthermore, in the event of reflection, approximately 50\% of the photons are absorbed by the light trap. As a result, we assume the reflectance of the entrance window, $R_e$, to be zero.

\subsection{The PMT reflectance}

As discussed earlier, the average reflectance of the PMT, $R_v$, is not zero due to the reflection in both the PMT window and the photocathode. The reflectance of the PMT window is determined only by its refractive index and as such it can be predicted using the Fresnel equations for dielectrics. As for the reflectance of the photocathode, we assumed it to be a constant $R_{ph}$ with no dependence on the angle of incidence. The average reflectance of the quartz window is, $R_q = \overline{\mathcal{F}}_{\nicefrac{n_{\mathrm{SiO_2}}}{n_1}}$ (see eq.\,\ref{IntegralFo}), where $n_{\mathrm{SiO_2}}$ is the refractive index of the fused quartz. By considering all the possible multiple reflections, we can arrive to:
\begin{equation}
R_v = R_q + \frac{R_{ph}\left(1-R_q\right)^2}{1-R_{ph} R_q}.
\end{equation}

\subsection{The hemispherical reflectances}
The first reflection inside the sphere occurs at the normal direction and corresponds to the hemispherical reflectance, denoted as $R_1$. The bi-hemispherical reflectance, denoted as $R$, gives the probability of reflection in the subsequent reflections.   
The factor $R_1$ corresponds to the hemispherical reflectance in the first reflection inside the sphere, which occurs at normal direction, and $R$ to the bi-hemispherical reflectance. Both factors are defined with the following integrals:
\begin{eqnarray}
R_1 & = &  \int_{2\pi} f_r \left(0, 0;\theta_r, \phi_r \right) \mathrm{d}\Omega_r,\\
R & = & \frac{1}{\pi} \int_{2\pi} \int_{2\pi} f_r \left(\theta_i, \phi_i;\theta_r, \phi_r \right) \mathrm{d}\Omega_r \mathrm{d} \Omega_i.
\end{eqnarray}
where $f_r$ corresponds to the BRIDF given by eq.\,\ref{BRIDF_mymodel} and the differentials are defined by eq.\,\ref{eq:dOmega}.

The result of both integrals is:
\begin{equation}
\begin{split}
R_1  = & F\left(0;\frac{n_2}{n_1}\right)\cdot E\left(\sigma_\alpha \right) + \\
  & + \frac{\rho}{1-\rho\overline{\mathcal{F}}_{\nicefrac{n_1}{n_2}}}\left[1-{F}\left(0;\frac{n_2}{n_1}\right)\right]\cdot\left[1-\overline{\mathcal{F}}_{\nicefrac{n_1}{n_2}}\right],\label{EqofR1}\\
 \end{split}
 \end{equation}
\begin{equation}
\begin{split}
R   = & \overline{\mathcal{F}}_{\nicefrac{n_2}{n_1}}+  \\ & + \frac{\rho}{1-\rho\overline{\mathcal{F}}_{\nicefrac{n_1}{n_2}}}\left[1-\overline{\mathcal{F}}_{\nicefrac{n_2}{n_1}}\right]\cdot\left[1-\overline{\mathcal{F}}_{\nicefrac{n_1}{n_2}}\right],
\end{split}
\end{equation}
%\end{align}
where the factor $E$ accounts for a fraction of the light specularly reflected in the first reflection in the east port and then exited through the entrance port. This factor is zero for smooth surfaces ($\sigma_\alpha=0$) and increases with the roughness of the surface being $0.6$ to $\sigma_\alpha=0.17$.

\subsection{The light absorption}

The factor $\eta_a$ quantifies the reduction in the throughput of the sphere due to light absorption. It is given by the product of the absorption coefficient $a_{\mathrm{abs}}$ and the average path length $d_{\mathrm{mean}}$ between two consecutive reflections inside the sphere, as follows:
\begin{equation}
\eta_a = d_{\mathrm{mean}} \cdot a_{\mathrm{abs}},
\label{eq:TISabs}
\end{equation}
To estimate $d_{\mathrm{mean}}$, we assume that the reflected photons follow a Lambertian distribution. The distance between two points in a sphere centered at $(0,0,r)$ is given by $2r\cos\theta$, where $\theta$ is the angle between the two points as measured from the sphere's center. Therefore, the average distance between two consecutive reflections inside the sphere is:
\begin{equation}
d_{\mathrm{mean}} = \frac{4r}{3}.
\end{equation}
Here, $r$ is the radius of the sphere. Note that the factor $\nicefrac{4}{3}$ is the average of the squared cosine of the angle between two points in a sphere with uniform radiance integrated over the hemisphere.

\subsection{The effect of the baffle and roughness}

To account for the effect of the baffle and the surface's roughness, we replaced the sphere's reflectance, $R$, in eq.\,\ref{eq:SphereEquation} with  $R^b$. The $b$ was obtained by adjusting this factor to the simulation output in the region $\rho\in[0.8, 0.99]$. It was determined to be 1.07 in the air and 1.12 in the liquid.

\section{Correction to the Incident Flux}\label{AppendixE}

The incident flux, $\Phi_I$, is given by the equation \ref{eq:incidentfluxcorrection}, which depends on the reflectance of the PMT window, $R_{v}$, and the reflectance of the entrance port, $R_{e}$. Since both the PMT window and the entrance window of the west port are formed by two interfaces, the reflectance accounts for multiple reflections resulting in a geometric series:
\begin{eqnarray}
%R_{\mathrm{PMT}} & = & R_{q,0}  + \frac{\left(1-R_{q,0}\right)^2}{1-R_{q,0}R_{ph}}R_{ph},\quad \mathrm{and}\\
%R_{\mathrm{W}} & = & R_{q,0}  + \frac{\left(1-R_{q,0}\right)^2}{1-R_{q,0}R_{e,0}} R_{e,0},
R_{v} & = & Q  + \frac{\left(1-Q\right)^2}{1-Q\cdot R_{ph}}R_{ph},\quad \mathrm{and}\\
R_{e} & = & Q  + \frac{\left(1-Q\right)^2}{1-Q\cdot V} V,
\label{CorrFluxes}
\end{eqnarray}
where $Q=F\left(0; \frac{n_\mathrm{SiO_2}}{n_1}\right)$  is the reflection probability between the air or liquid inside the sphere and the fused quartz window of the PMT or the entrance port. $V=F\left(0; n_{\mathrm{SiO_2}}\right)$ is the reflection probability between the fused quartz window of the entrance port and the outside of the sphere, which is air.

\section*{Acknowledgements}

This work was supported by the Portuguese Foundation for Science and Technology (FCT) under the award numbers PTDC/FIS-PAR/2831/2020, and POCI/01-0145-FEDER-029147, PTDC/FIS-PAR/29147/2017, funded by OE/FCT, Lisboa2020, Compete2020, Portugal2020, FEDER. 
Cláudio Silva was supported by the IF/00877/2015 funded by FCT.

\bibliographystyle{apsrev4-2}
\bibliography{main}% Produces the bibliography via BibTeX.

%\end{document}

%\input{Supplemental}

\end{document}